\documentclass[aps,prd,preprint,superscriptaddress,nofootinbib,amsmath,amssymb]{revtex4}
\pdfoutput=1
\usepackage{graphicx}
\usepackage{dcolumn}
\usepackage{bm}
\usepackage{amsmath}
\usepackage{amsfonts}
\usepackage{amssymb}
\usepackage{appendix}
\usepackage{pdfpages}
\usepackage{graphicx}
\usepackage{wrapfig}
\usepackage{multirow}
\usepackage{mathrsfs}
\usepackage{epstopdf}
\usepackage{slashed}
\usepackage{soul}
\usepackage{mathtools}
\usepackage{floatrow}
\usepackage{float}
\usepackage{csquotes}
\usepackage{wrapfig}
\usepackage{graphicx}
\usepackage{epsfig}
\usepackage[utf8]{inputenc}
\usepackage{epsfig}
\usepackage{dcolumn}
\usepackage{morefloats}
\usepackage{hyperref}
\usepackage{stackengine,filecontents,amsmath}
\newcommand{\bea}{\begin{eqnarray}}
\newcommand{\eea}{\end{eqnarray}}
\newcommand{\MoreRep}[2]{\mbox{\textbf{#1}} ^{\textbf{#2}}}
\newcommand{\Rep}[1]{\mbox{\textbf{#1}}}

\graphicspath{{figs/}}

\catcode`\@=11
\def\lsim{\mathrel{\mathpalette\@versim<}}
\def\gsim{\mathrel{\mathpalette\@versim>}}
\def\@versim#1#2{\vcenter{\offinterlineskip
\ialign{$\m@th#1\hfil##\hfil$\crcr#2\crcr\sim\crcr }}}
\catcode`\@=12

\newcommand{\AddrUNAM}{Instituto de Física, Universidad Nacional Autónoma de México, A.P. 20-364, Ciudad de México 01000, México.}
\newcommand{\AddrUCN}{Departamento de Física, Universidad Católica del Norte, Avenida Angamos 0610, Casilla 1280, Antofagasta, Chile.}

\begin{document}
\title{ Neutrino phenomenology in a left-right $D_4$ symmetric model }

\author{Cesar Bonilla}\email{cesar.bonilla@ucn.cl}\affiliation{\AddrUCN}
\author{Leon M.G. de la Vega}\email{leonm@estudiantes.fisica.unam.mx}\affiliation{\AddrUNAM}
\author{R. Ferro-Hernandez}\email{ferrohr@estudiantes.fisica.unam.mx}\affiliation{\AddrUNAM}
\author{Newton Nath}\email{newton@fisica.unam.mx}\affiliation{\AddrUNAM}
\author{Eduardo Peinado} \email{epeinado@fisica.unam.mx}\affiliation{\AddrUNAM}


\begin{abstract}

%

We present a minimal left-right  symmetric flavor model and analyze 
the predictions for the neutrino sector. In this
scenario, the Yukawa sector is shaped by the dihedral $D_4$ symmetry which leads
to correlations for the neutrino mixing parameters. We end up with the four possible solutions within this model. 
We further analyzed  the impact of  the  upcoming long-baseline neutrino oscillation experiment,  DUNE. Due to its high sensitivity, it has been observed that the DUNE will be  able to rule out two  of the solutions. Finally, the predictions for the neutrinoless double beta decay and the lepton flavor violating process $\mu \rightarrow e \gamma$ for the model have also been examined.
%
\end{abstract}

\maketitle

\section{Introduction}

A number of phenomenal experimental evidences over the past two decades have established the fact that neutrinos oscillate through their propagation path ~\cite{Abe:2011sj,Adamson:2011qu,An:2012eh,Ahn:2012nd}, which implies non-zero neutrino masses and mixings.  This fact provides an undoubtedly motivation for the existence of physics beyond the Standard Model (SM), as neutrinos are massless in the SM. Furthermore, the experimental efforts in understanding the neutrino properties have determined the two mass-squared 
differences and three sufficiently large leptonic mixing angles. From the global analysis  of neutrino oscillation  data~\cite{deSalas:2017kay} (other global analysis can be found in 
\cite{Capozzi:2016rtj,Esteban:2018azc}), the best fit values  
and the $1\sigma$ intervals for a normal neutrino mass ordering (NO) are
given by \footnote{Note here that the neutrino oscillation experiments are sensitive to (mass)$ ^{2} $- differences and hence, the possibility of a massless neutrino is not excluded.}
\begin{eqnarray}
& &|\Delta m_\text{sol}^2|= 7.55^{+0.20}_{-0.16}\, \times 10^{-5}\,\text{eV}^2,\ \ |\Delta m_\text{atm}^2|= 2.50\pm 0.03\, \times 10^{-3}\,\text{eV}^2, \notag \\
& & \theta_{12} / ^{\circ} =34.5^{+1.2}_{-1.0}, \ \
\theta_{13} / ^{\circ} =8.45^{+0.16}_{-0.14}, \ \ \theta_{23} / ^{\circ} =47.7^{+1.2}_{-1.7},
\ \ \text{and}\ \ \delta_{\text{CP}} / ^{\circ} =218^{+38}_{-27} \;. \ \
\end{eqnarray}
Moreover, the theory behind the dynamical
origin of the  masses of neutrinos along with their flavor mixing patterns and whether they are Majorana or Dirac
fermions, is yet unanswered. The simplest idea behind these shortcomings relies on the assumption that the neutrinos are Majorana particles and 
their tiny masses are generated through a seesaw mechanism~\cite{Minkowski:1977sc, Yanagida:1979as, Mohapatra:1979ia, Schechter:1980gr, Schechter:1981cv, Foot:1988aq}.
Interesting extensions of the SM featured by the inherent new physics 
signatures are those that consider a left-right (LR) symmetric nature \cite{Pati:1974yy,Mohapatra:1974gc,Mohapatra:1974hk,Senjanovic:1975rk}.
For instance, the LR symmetric models have the virtue of accounting for the small neutrino 
masses from the contribution of two mechanisms, the type-I and type-II seesaw,
which implies the existence of new particles.

The simplest LR symmetric model is dictated by the gauge symmetry group $\text{SU}(3)_C \otimes \text{SU}(2)_L \otimes \text{SU}(2)_R \otimes\text{ U}(1)_{B-L}$. In this case, the 
fermion fields   have the following charge assignments~\cite{Deshpande:1990ip}, 
\begin{equation}
l_{L}\simeq {\bf(1,2,1,-1)},\ \  
l_{R}\simeq {\bf(1,1,2,-1)},\ \ 
Q_{L}\simeq {\bf(3,2,1,1/3)}\ \ \text{and} \ \
Q_{R}\simeq {\bf(3,1,2,1/3)},
\end{equation}
whereas the scalar potential is formed by the two triplets and one bi-doublet
whose LR charges are,
\begin{equation}
\Delta_{L}\simeq {\bf(1,3,1,2)},\ \ 
\Delta_{R}\simeq {\bf(1,1,3,2)}\ \ \text{and} \ \
\Phi       \simeq {\bf(1,2,2,0)}.
\end{equation}
If the LR breaking scale and the masses of the new scalar fields
are of $\mathcal{O}(\text{TeV})$, this minimal setup produces 
sizeable contributions to the lepton flavor violating (LFV) decays,
lepton number violation as well as CP violating processes~\cite{Hirsch:1996qw,Zhang:2007da,Tello:2010am,Awasthi:2013ff,Barry:2013xxa,Awasthi:2015ota,Bambhaniya:2015ipg}. 
Therefore, this scenario turns out to be very appealing for
the experimental searches among the low-energy LFV processes~\cite{Bonilla:2016fqd}.
Further constraints apply to this model from the LHC searches of new physics~\cite{Chen:2013fna,Bambhaniya:2013wza,Mohapatra:2013cia,Bambhaniya:2015wna,Dev:2016dja}. On the other hand, the LR symmetry is also possible to be broken at 
higher energies, such  as the grand unification theory (GUT) scale,
leading to the gauge coupling unification~\cite{Chang:1983fu,Chang:1984uy}.
This makes the LR models interesting frameworks from the perspective of the GUTs like 
$SO(10)$~\cite{Fritzsch:1974nn, Arbelaez:2013nga}.

On top of gauge symmetries, one can impose additional global symmetries that relate the flavor structure of the SM. In the past decade, there have been numerous amount of works in this direction, for reviews see \cite{Ishimori:2010au,Altarelli:2010gt}. Nevertheless, it is particularly interesting to examine the interplay between the LR symmetry and a discrete flavor symmetry. This combination  shapes and correlates the Yukawa sector, giving  predictions for the flavor observables, i.e., neutrino masses and mixings ~\cite{Rodejohann:2015hka,Gomez-Izquierdo:2017rxi,Das:2018rdf,CarcamoHernandez:2018hst,Garces:2018nar}. In this work, we study the effects  of combining a non-Abelian discrete flavor symmetry group $D_4$ with the  LR symmetry. It is worthwhile to mention here that the $D_4$ flavor symmetry group has been explored in literature in \cite{Grimus:2003kq,Grimus:2004rj,Babu:2004tn,Ko:2007dz,Adulpravitchai:2008yp,Das:2019itj,CarcamoHernandez:2020ney,Ishimori:2008gp,Ishimori:2008ns,Hagedorn:2010mq,Meloni:2011cc,Vien:2013zra,AhlLaamara:2017gsd,Kobayashi:2018zpq},  not in combination with a LR symmetric model to the best of our knowledge.
It is to be noted further that a $Z_2$ symmetry has also been augmented with the $D_4$ flavor group in the model. We like to stress here that our intention is to keep the left-right scalar sector as the one in the Standard LR model, namely a bi-doublet and  two triplets (left and right).
Hence, in order to break the flavor symmetry we include  the flavon fields that are charged under $Z_2$ symmetry as well as the bi-doublet which  also carries $Z_2$ charge to obtain the desired Lagrangian in this model.
 Among many of the consequences of this model, a noteworthy outcomes of the model is the appearance of the two-zero texture of the neutrino mass matrix. 
Under the Glashow-Frampton-Marfatia classification \cite{Frampton:2002yf} for the two-zero texture Majorana neutrino mass matrices, we get an $A_2$ type texture zero matrix. This model also predicts a non-diagonal mass matrix for the charged leptons. 

The outline of the paper is as follows: in Sec.~\ref{sec:model} we present the model and the necessary charge  assignments. In this section, we also give the invariant Lagrangian of the theory and derive the leptonic mass matrices. We explain the procedure of our analysis in Sec.~\ref{sec:NuPheno} as well as
show our results for the neutrino predictions within the model. Our final
comments and summary are given in Sec.~\ref{sec:conclusions}.  Appendix~\ref{sec:d4multiplication} summarizes the $D_4$ algebra, whereas the necessary equations for the LFV processes are given in appendix~\ref{AppendixLFV}.

\section{Left-right $D_4$ Symmetric Model}
\label{sec:model}

We consider an extension of the minimal left-right symmetric model by adding a $D_4$ flavor symmetry. Besides postulating a symmetry that shapes the Yukawa sector, we add two flavon fields, $\chi$ and $\eta$
transforming as a singlet and doublet under $D_4$, respectively. In Table~\ref{Table1}
we provide the matter content and charge assignments of the model. In this framework, the 
symmetry breaking goes like
\begin{equation}
\text{LRSM}\otimes G_\text{F} \overset{\eta,~\chi~~}{\longrightarrow}\text{LRSM}\overset{\Delta}{\longrightarrow}\text{SM}\overset{\Phi}{\longrightarrow} \text{SU}(3)_C\otimes\text{U}\left(1\right)_{em}\notag,
\end{equation}where $G_F=D_4\otimes Z_2$ and its breaking is associated to the non-zero
vacuum expectation values (vevs) of the flavon fields $\langle \chi\rangle$ and $\langle \eta\rangle$.

 \begin{table}[h]
\centering{}
\begin{tabular}{|c||c|c||c|c|c||c|c|}
\hline 
 & $\ell_{{L}_{D(S)}}$ & $\ell_{{R}_{D(S)}}$ & $\Delta_{L}$ & $\Delta_{R}$  & $\Phi$ & $\eta$ & $\chi$\tabularnewline
\hline 
\hline 
$\text{SU}\left(2\right)_{L}$ & \textbf{2} & \textbf{1} & \textbf{3} & \textbf{1} & \textbf{2} & \textbf{1} & \textbf{1}\tabularnewline
\hline 
$\text{SU}\left(2\right)_{R}$ & \textbf{1} & \textbf{2} &  \textbf{1} & \textbf{3} & \textbf{2} &\textbf{1} & \textbf{1}\tabularnewline
\hline 
$\text{U}\left(1\right)_{B-L}$ & -1 & -1 & 2 & 2 & 0 & 0 & 0\tabularnewline
\hline 
$D_{4}$ & \textbf{2$\oplus$1} & \textbf{2$\oplus$1} & \textbf{1} & \textbf{1} & \textbf{1} &  \textbf{2}& \textbf{1}\tabularnewline
\hline 
$Z_{2}$ & 1 & 1  & 1 & 1 & -1 & -1 & -1\tabularnewline
\hline 
\end{tabular}\caption{\footnotesize Matter content and charge assignments of the left-right $D_{4}$ model, where  $D^{} (S)$ stands for $D_{4}$ doublet (singlet). Three lepton families are arranged in  a doublet and a singlet of $D_{4}$, respectively. Notice also that only the SM Higgs doublet $\Phi$ and the  flavon fields $\eta, \chi$ are kept charged under the $Z_2$ symmetry.
\label{Table1}}
\end{table}
%
%
 We assume the following sequential symmetry breaking 
$\Lambda_\text{F} >> \Lambda_\text{LR}>>\Lambda_\text{EW}$, where $\Lambda_F$ is the flavour breaking scale and $\Lambda_\text{LR}$ is the left-right symmetry breaking scale~\footnote{With this assumption the 
flavon fields decouple from the theory having only an impact on the Yukawa couplings. Then, 
in this energy regime the scalar potential is approximate  to the  minimal LRSM one~\cite{Deshpande:1990ip}. 
A detailed discussion of the LFV process in this model has been presented in the next section.
}.

Given the matter content shown in Table~\ref{Table1}, the Yukawa 
Lagrangian (up to dimension-5) for the leptons can be expressed as
\begin{eqnarray}
\label{LY}
\mathcal{L}_Y
& \supset & \bar{\ell}_{L_{D}}\left(y_{1}\frac{\chi}{\Lambda_F}\Phi+\tilde{y}_{1}\frac{\chi}{\Lambda_F}\tilde{\Phi}\right)\ell_{R_{D}}+\bar{\ell}_{L_{D}}\left(y_{2}\frac{\eta}{\Lambda_F}\Phi+\tilde{y}_{2}\frac{\eta}{\Lambda_F}\tilde{\Phi}\right)\ell_{R_{s}} \notag\\
 & + & \bar{\ell}_{L_{s}}\left(y_{3}\frac{\eta}{\Lambda_F}\Phi+\tilde{y}_{3}\frac{\eta}{\Lambda_F}\tilde{\Phi}\right)\ell_{R_{D}}+\bar{\ell}_{L_{s}}\left(y_{4}\frac{\chi}{\Lambda_F}\Phi+\tilde{y}_{4}\frac{\chi}{\Lambda_F}\tilde{\Phi}\right)\ell_{R_{s}}\notag\\
 & + & \frac{Y_{L_{1}}}{2}\ell_{L_{D}}^{T}C\left(i\sigma_{2}\right)\Delta_{L}\ell_{L_{D}}+\frac{Y_{L_{2}}}{2}\ell_{L_{s}}^{T}C\left(i\sigma_{2}\right)\Delta_{L}\ell_{L_{s}}\notag\\
 & + & \frac{Y_{R_{1}}}{2}\ell_{R_D}^{T}C\left(i\sigma_{2}\right)\Delta_{R}\ell_{R_{D}}+\frac{Y_{R_{2}}}{2}\ell_{R_{s}}^{T}C\left(i\sigma_{2}\right)\Delta_{R}\ell_{R_{s}}+\text{h.c.}
\end{eqnarray}
where $\ell_{L_{D}} = (\ell_{L_1}, \ell_{L_2})^T$, $(\ell_{R_{D}} = (\ell_{R_1}, \ell_{R_2})^T)$ represents left (right)-handed $D_4$ doublet, $\ell_{L_{s}} = \ell_{L_3}$ ($\ell_{R_{s}} = \ell_{R_3}$) left (right)-handed $D_4$ singlet and the bi-doublet $\Phi$ can be read as
\begin{equation}
 \Phi = \begin{pmatrix}
\phi_1^0 & \phi_1^+ \\ 
\phi_2^- & \phi_2^0
\end{pmatrix}.
\end{equation}
In appendix \ref{sec:d4multiplication} we give a detailed discussion of the $D_4$ algebra that has been utilized to derive the Lagrangian as given in Eq.~(\ref{LY}).
An explicit Lagrangian has also been presented using the $D_4$ product rules that are used to derive Eq.~(\ref{LY}).
Moreover, it is worth mentioning that the other two possible choices to assign the matter fields namely,  doublets and singlets of $D_4$ are equivalent  to the one in Table~\ref{Table1} and a detailed discussion has been given in  appendix~\ref{sec:d4multiplication}.

Note that the Dirac lepton mass matrices stem from the dimension-5 operators. Hence, from Eq.~(\ref{LY}) after the spontaneous 
symmetry breaking (SSB), one gets that the mass matrix for the charged leptons as
\begin{eqnarray}
M_{\ell}=\frac{1}{\sqrt{2}}(Y'_{L}v_2 + \tilde{Y}^{'}_L v_1) \;,
\end{eqnarray}
where
\bea 
Y'_L =\frac{1}{\Lambda_F}\left(
\begin{array}{ccc}
   0            &   y_1 v_\chi    & y_2 v_{\eta_2}\\
y_1 v_\chi      &   0      & y_2 v_{\eta_1} \\
y_3 v_{\eta_2}  &   y_3 v_{\eta_1}      & y_4 v_\chi
\end{array}\right)\ \
\text{and}
\ \
\tilde{Y}'_{L} =\frac{1}{\Lambda_F}\left(
\begin{array}{ccc}
   0            &   \tilde{y}_1 v_\chi    & \tilde{y}_2 v_{\eta_2}\\
\tilde{y}_1 v_\chi      &   0      & \tilde{y}_2 v_{\eta_1} \\
\tilde{y}_3 v_{\eta_2}  &   \tilde{y}_3 v_{\eta_1}      & \tilde{y}_4 v_\chi
\end{array}\right),
\eea
with $\langle \chi \rangle\equiv v_\chi $, $\langle\eta\rangle \equiv (v_{\eta_1},v_{\eta_2})^T$ and $\Phi$ 
\begin{equation}
\langle \Phi \rangle =  \begin{pmatrix}
v_1 & 0 \\ 
0 & v_2
\end{pmatrix} .
\end{equation}
Assuming a vev alignment $\langle \eta \rangle\sim(1,0)^T$, the mass matrix for the charged leptons becomes
\bea
M_{\ell} =
\label{MMCL}
\frac{1}{\sqrt{2} \Lambda_F}\left(
\begin{array}{ccc}
   0                                              &   (y_1  v_{2}+ \tilde{y}_1 v_1)v_\chi & 0\\
  (y_1  v_{2}+ \tilde{y}_1 v_1)v_\chi             &   0                                  & (y_2  v_{2}+ \tilde{y}_2 v_1)v_{\eta_1} \\
   0                                              &   (y_3  v_{2}+ \tilde{y}_3 v_1)v_{\eta_1} & (y_4  v_{2}+ \tilde{y}_4 v_1)v_{\chi}
\end{array}\right) \;.
\eea
The matrix $M_{\ell}$
can be diagonalised by a bi-unitary transformation as
\begin{equation}
\label{UMVdagger}
\text{diag}(m_e,m_\mu,m_\tau)=U_{\ell}M_{\ell}V_{\ell}^\dagger \;,
\end{equation}
and the neutrino mass matrix  is given  by
\bea
m_{\nu} =\left(
\begin{array}{cc}
   m_L     &   m_D \\
   m_D     &   m_R
\end{array}\right) \;,
\eea
where $m_L=\sqrt{2}Y_L v_L$ and $m_R=\sqrt{2}Y_R v_R$,
with  
\bea
\label{YLR}
Y_{L(R)} =\left(
\begin{array}{ccc}
   0     &   Y_{L_1(R_1)} & 0\\
   Y_{L_1(R_1)}     &   0        & 0 \\
   0     &   0        & Y_{L_2(R_2)}
\end{array}\right).
\eea
In this scenario, the Dirac neutrino mass matrix turns out to be
\begin{equation}
\label{mDnu}
 m_D= \frac{1}{\sqrt{2}}(Y'_{L}v_1 + \tilde{Y}^{'}_L v_2) \;.
\end{equation}
After the SSB, the light neutrino eigenstates acquire their masses
through the type-I and type-II seesaw mechanism. Hence, the light-neutrino mass matrix is given by,
%
\begin{equation}
\label{MMnu}
M_\nu^{\text{light}}=m_L-m_D m_R^{-1} m_D^T \;,
\end{equation}
where $v_R>> v_L, v_1, v_2$ has been assumed.

The left-right symmetric nature of the theory demands a relation between the Yukawa couplings mediating the interaction of leptons with the scalar
triplets, i.e. $Y_R=Y_L$.
The left-right exchange symmetry can be realized through either C or P transformations. Here we choose to use P-transformations, which demand the hermiticity of Dirac type fermion mass matrices, that is,
\begin{equation}
M_{\ell}= M_{\ell}^\dagger\;, \quad m_D=m_D^\dagger \;.
\end{equation}

\section{Neutrino Phenomenology}
\label{sec:NuPheno}

Having discussed the theoretical framework in the previous section, here we aim to discuss various phenomenological  importance of the model.
In doing so, we notice from Eq.~(\ref{MMCL})  that the mass matrix for charged leptons is 
non-diagonal. The left-right symmetry gives further relations for leptonic 
Yukawas, as mentioned in the previous section. Using this fact, the mass 
matrix for charged leptons, as given by Eq.~(\ref{MMCL}), can be recasted as
\begin{equation}
\label{MCLorth}
M_{\ell}=\left(\begin{array}{ccc}
0 & a_{\ell} & 0\\
a_{\ell}^* & 0 & b_{\ell}\\
0 & b_{\ell}^* & c_{\ell}
\end{array}\right) \;.
\end{equation}
The phases  of this matrix can be absorbed in a pair of diagonal phase matrices ($P$ and $P'$), this will lead to a real charged-lepton matrix as
\begin{equation}\label{eq:ChargedLepton1}
\tilde{M}_{\ell}=P M_{\ell} P'.
\end{equation}
In this basis the neutrino mass matrix becomes
\begin{equation}\label{eq:MnuTilde}
\tilde{M}_\nu = P^T M_\nu P,
\end{equation}
where $M_\nu$ is the neutrino mass matrix in the interaction basis as given by Eq. (\ref{MMnu}).
Since $\tilde{M}_\ell$ is symmetric, it can be diagonalised as
 \begin{equation}\label{eq:DiagCL}
\text{diag}(m_e,m_\mu,m_\tau)=O_{\ell}\tilde{M}_{\ell}O_{\ell}^T\notag \;,
\end{equation}
where $O_\ell$ is an orthogonal matrix and one can easily
get the expressions for $a_\ell\, , b_\ell$ and $c_\ell$ in terms of the charged-lepton masses. This is done by computing the invariants of the charged-lepton mass matrix, 
namely $\text{Tr}(M_\ell)$, $\text{Tr}(M_\ell^2)$ and $\text{det}(M_\ell)$~ 
\footnote{Notice that the $\text{det}(M_\ell)$ is negative and hence one of the masses carries negative sign. Here we choose $m_{\mu} = - |m_{\mu}|$.}. Then, the 
matrix elements in Eq.~(\ref{MCLorth}) as functions of the masses can be read as
\begin{eqnarray}
a_{\ell}&=&\pm\frac{\sqrt{m_{e}m_{\mu}m_{\tau}}}{\sqrt{m_{e}-m_{\mu}+m_{\tau}}} \;, \notag\\
b_{\ell}&=&\pm\frac{\sqrt{-m_{\mu}+m_{\tau}}\sqrt{-m_{e}^{2}+m_{e}m_{\mu}-m_{e}m_{\tau}+m_{\mu}m_{\tau}}}{\sqrt{m_{e}-m_{\mu}+m_{\tau}}} \;,
\notag \\
c_{\ell}&=&m_{e}-m_{\mu}+m_{\tau} \;.
\end{eqnarray}
With this information one can compute the rotation matrix $O_{\ell}$ as
\begin{equation}\label{eq:U-ChargedLepton}
O_{\ell}=\left(\begin{array}{ccc}
0.998 & -\text{sgn}(a_{\ell})\,0.070 & \text{sgn}(a_{\ell} b_{\ell})\,0.001\\
\text{sgn}(a_{\ell})\, 0.068 & 0.969 & \text{sgn}(b_{\ell})\,0.236\\
-\text{sgn}(a_{\ell} b_{\ell})\,0.017 & -\text{sgn}(b_{\ell})\,0.235 & 0.972
\end{array}\right) \;.
\end{equation}
Note that $O_\ell$ in Eq.~(\ref{eq:U-ChargedLepton}) is determined up to sign combinations
of the parameters $a_\ell$ and $b_\ell$.\\

Regarding the neutrino mass matrix, this is obtained by using Eqs.~(\ref{YLR}-\ref{MMnu}) and
turns out to be,
\begin{equation}
\label{Mnu}
\tilde{M}_{\nu}=\left(\begin{array}{ccc}
0 & a_{\nu} & 0\\
a_{\nu} & d_{\nu} & b_{\nu}\\
0 & b_{\nu} & c_{\nu}
\end{array}\right)\;,
\end{equation}
where $a_{\nu}$, $b_{\nu}$, $c_{\nu}$, and $d_{\nu}$ are complex entries. Therefore, the diagonalization
of $\tilde{M}_\nu$ as given by Eq.~(\ref{Mnu}) leads to the active light neutrino masses.
%
 This mass matrix and the neutrino mass 
eigenstates are related as follows,
\begin{equation}
\label{eq:leptonmixing}
 \tilde{M}_\nu= U_\nu^*(\theta^0_{12},\theta^0_{23},\theta^0_{13},\delta_0)\, \text{diag}(m_{\nu_1}, m_{\nu_2}, m_{\nu_3}) \,U_\nu^{\dagger}(\theta^0_{12},\theta^0_{23},\theta^0_{13},\delta_0) \;, 
\end{equation}
where $m_{\nu_i} $ are the light neutrino masses and the unitary matrix $U_\nu$ follows the PDG 
parameterization~\cite{Tanabashi:2018oca}. Therefore, in this model, the lepton mixing matrix (also known as Pontecorvo-Maki-Nakagawa-Sakata (PMNS) mixing matrix~ \cite{MNS,Pontecorvo:1957qd}) is defined by~\footnote{Similar structure for charged leptons and neutrinos was obtained in the context of $S_3$ flavor symmetry~\cite{Meloni:2010aw,Meloni:2012sy}.},
\begin{equation} 
\label{eq:pmns_sol}
V_{L}(\theta_{12},\theta_{23},\theta_{13},\delta)= O_{\ell}^{T}U_{\nu}(\theta^0_{12},\theta^0_{23},\theta^0_{13},\delta_0) K \;,
\end{equation}
where the angles $\theta_{ij}~ (i < j = 1, 2, 3)$ correspond to the mixing angles determined by neutrino oscillation experiments, 
$\delta$ represents the Dirac CP-violating phase and $K$ is the diagonal Majorana phase  matrix.

\subsection{Results}
\begin{figure}[t]
\begin{centering}
\includegraphics[scale=0.16]{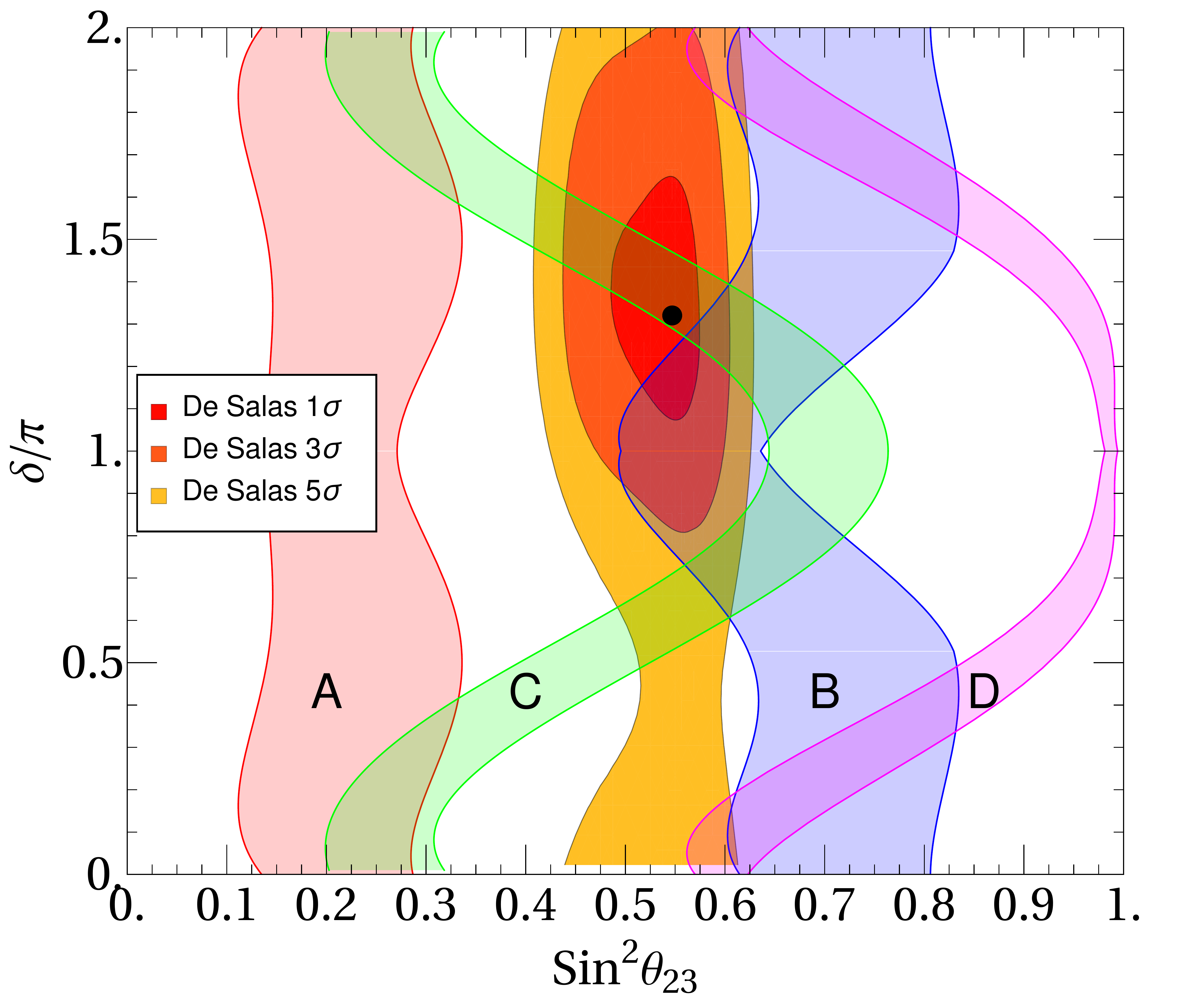}
\includegraphics[scale=0.16]{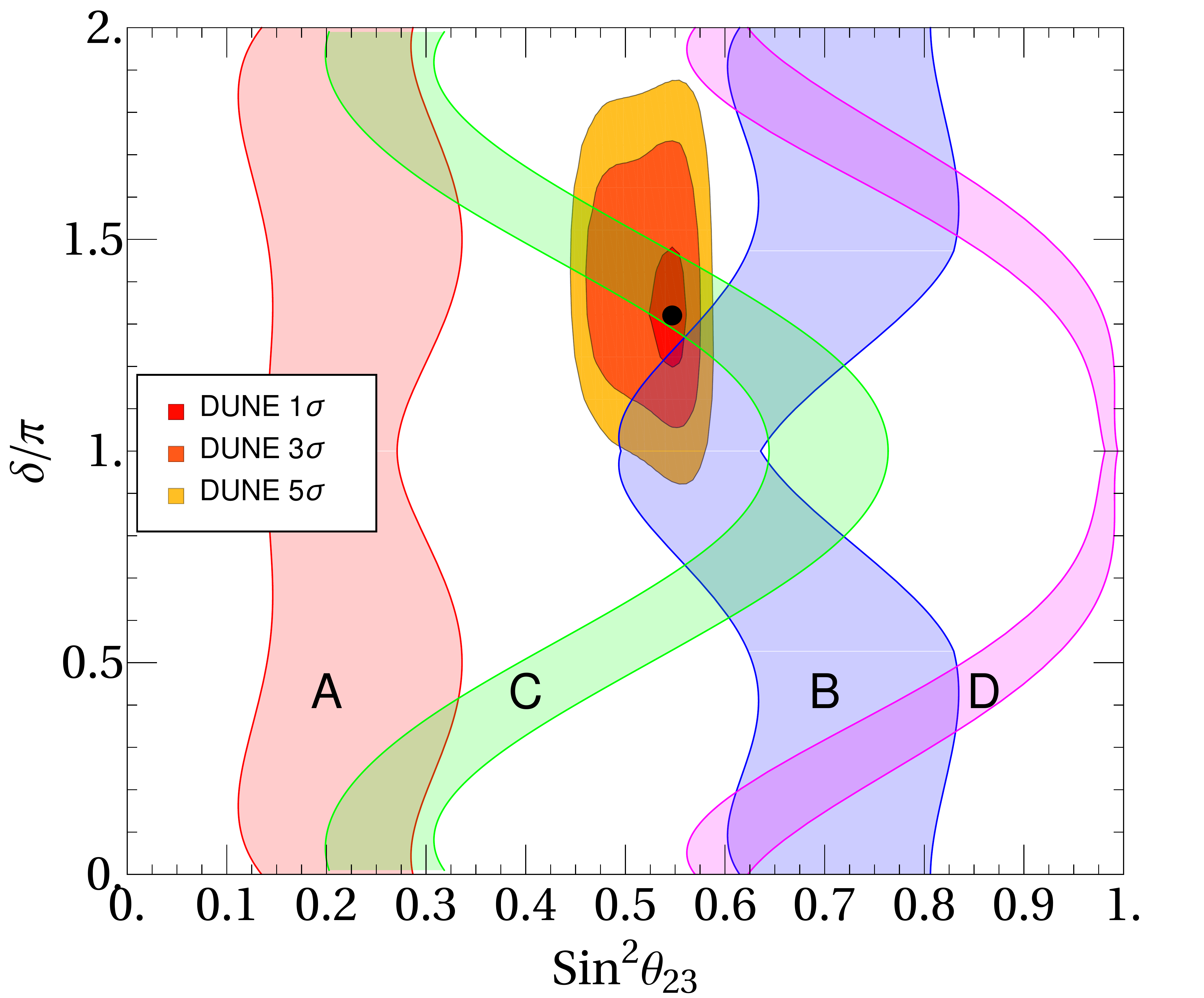}
\par\end{centering}
\begin{centering}
\caption{\footnotesize Allowed parameter space in ($\sin^{2}\theta_{23} - \delta$) plane for the four-solutions. Various colors viz, light-red, light-blue, light-green, and light-pink show correlation for A = (+, -), B = (-, -), C = (+, +), and D = (-, +), respectively. The solid contours for the left (right)-panel  depicts the allowed region for ``global-fit data" (``simulated results of DUNE") corresponding the latest best-fit value as shown by black-dot.}
\label{fig:Th23-CP}
\end{centering}
\end{figure}

According to the latest global analysis of the neutrino oscillation data~\cite{deSalas:2017kay,Capozzi:2016rtj,Esteban:2018azc}, among the six neutrino oscillation parameters, the atmospheric mixing angle $\theta_{23}$ and the Dirac CP-phase $\delta$ are the two least known parameters. 
Hence,  we show results of our numerical scan in the ($\sin^{2}\theta_{23} - \delta$) plane in Fig.~\ref{fig:Th23-CP}.
While  performing our numerical scan, we have varied the unphysical mixing angles $\theta^0_{ij}$ in the range $(0, \pi/2)$, whereas the unphysical Dirac CP phase $\delta_0$ has been scanned  over the range $(0, 2\pi)$ of matrix $U_\nu$ (see Eq.~\ref{eq:pmns_sol})) for the $A_2$ texture.
Moreover, 
the neutrino mass squared differences as computed by the global analysis of the neutrino oscillation data~\cite{deSalas:2017kay} have also been utilized.
Then we obtain the physical lepton mixing angles using Eq. (\ref{eq:pmns_sol}), and compare with their experimentally allowed values from the global fit. 
Note that depending on the sign of ($a_l, b_l$), see Eq.~(\ref{eq:U-ChargedLepton}), there are four possible 
solutions which correlate the atmospheric angle $ \theta_{23} $ and the Dirac type CP-violating phase $ \delta $. These are denoted as: in light-red $\text{A} = (+, -)$; 
in light-blue $\text{B} = (-, -)$; in light-green $\text{C} = (+, +)$; and, in light-pink $\text{D} = (-, +)$.
In the left-panel, the ($\sin^{2}\theta_{23} - \delta$) plane depicts the allowed regions considering 
the latest global analysis of neutrino oscillation data~\cite{deSalas:2017kay} at $1$, $3$, and $5\sigma$ C. L., respectively. 
These contours are shown using the red, orange, and yellow colors, respectively. The best-fit value has been marked with a `black-dot'.
It can be seen from the left-panel that the solution A is ruled out by the present data at $5 \sigma$ C. L., whereas 
the  solution D is marginally allowed at $3 \sigma$ C. L., but only for the CP-conserving values, 
namely around  $\delta = 0, 2\pi$. We also notice that the solutions B and C are allowed  at  $1 \sigma$ C. L. 
Furthermore, it can be seen  that among the four cases only the solution C is able to explain the latest best-fit value of neutrino oscillation data.   

Similarly, in the right-panel of Fig.~\ref{fig:Th23-CP}, we show the compatibility of the  model by considering the simulated results of the next generation long baseline oscillation experiment, DUNE~\cite{Acciarri:2016crz}. The allowed parameter space of DUNE 
in the ($\sin^{2}\theta_{23} - \delta$) plane is found using the latest best-fit value of neutrino oscillation data.
For the numerical simulation of DUNE, the \texttt{GLoBES} package was used \cite{Huber:2004ka, Huber:2007ji} along with the auxiliary files in Ref.~\cite{Alion:2016uaj}. A running time of 3.5 years was assumed in both neutrino and antineutrino modes for DUNE, i.e. DUNE[3.5 + 3.5]. The detailed numerical procedures that have been followed to simulate data coincides with the one performed 
in ~\cite{Nath:2018xkz,Nath:2018fvw}.
Notice from the right-panel that DUNE results would significantly improve the precision of both the parameters. It is observed 
that $\sin^{2}\theta_{23}$ is constrained to values between $(0.45, 0.58)$, whereas $\delta$ is restricted to the 
range $(0.95, 1.88)\pi$ at  $5 \sigma$ C. L. after DUNE[3.5 + 3.5] running time.
Therefore, one can infer that the precise measurement of both parameters ($\theta_{23}$ and $\delta$) by DUNE,
the solution D will be ruled out at $5 \sigma$ C. L., still allowed by the latest global-fit data.


In this model we also have a prediction for the lepton number violating processes such as the neutrinoless 
double beta decay ($ 0\nu\beta\beta $). %
Ongoing experiments that are looking for the signatures of $ 0\nu\beta\beta$ decays are namely, GERDA 
Phase-II~\cite{Agostini:2018tnm}, CUORE \cite{Alduino:2017ehq}, SuperNEMO \cite{Barabash:2011aa}, KamLAND-Zen \cite{KamLAND-Zen:2016pfg} and EXO \cite{Agostini:2017jim}. 
The half-life of these processes can be expressed as \cite{Rodejohann:2011mu,*Dev:2013vxa},
\begin{equation}
(T^{0\nu}_{1/2})^{-1} = G_{0\nu}|M_{0\nu}(A,Z)|^{2} |\langle m \rangle_{ee}|^{2} \;,
\label{eq:halflifedoublebeta}
\end{equation}
where $  G_{0\nu}$ represents the two-body phase-space factor, $ M_{0\nu} $ is the nuclear matrix element and 
$|\langle m \rangle_{ee}|$ is the effective Majorana neutrino mass.
The expression of $|\langle m \rangle_{ee}|$ is given by,
\begin{equation}
\label{0nubb}
|\langle  m_{ee} \rangle| = \left| \sum^3_{i = 1} m_i V^2_{L_{e i}} \right| \;,
\end{equation}
where $V_L$ stands for lepton mixing matrix as mentioned in Eq.~(\ref{eq:leptonmixing}).
We like to point out here that in the diagonal charged lepton basis, the $A_2$ type two zero texture for the Majorana neutrino mass matrix  gives $(M_{\nu})_{11} = 0 = (M_{\nu})_{13}$. Hence, leads to zero predictions for the effective Majorana neutrino mass $|\langle m \rangle_{ee}|$ as discussed in Ref.~\cite{Frampton:2002yf}, where the PMNS mixing matrix has been completely determined by the neutrino sector.
However, in our formalism the charged-lepton mass matrix is non-diagonal (see Eq.~(\ref{MCLorth})), while the  Majorana neutrino mass matrix  respects $A_2$ type two zero  texture as given by Eq.~(\ref{Mnu}).
Thus, one can notice that  contribution to the PMNS mixing matrix arises from both the leptonic mass matrices as given by Eq.(\ref{eq:pmns_sol}). Therefore, we end-up with non-zero predictions for the effective Majorana neutrino mass in this study as given by Eq.~(\ref{0nubb}).

\begin{figure}[H]
\begin{centering}
\includegraphics[height = 6cm, width = 8cm]{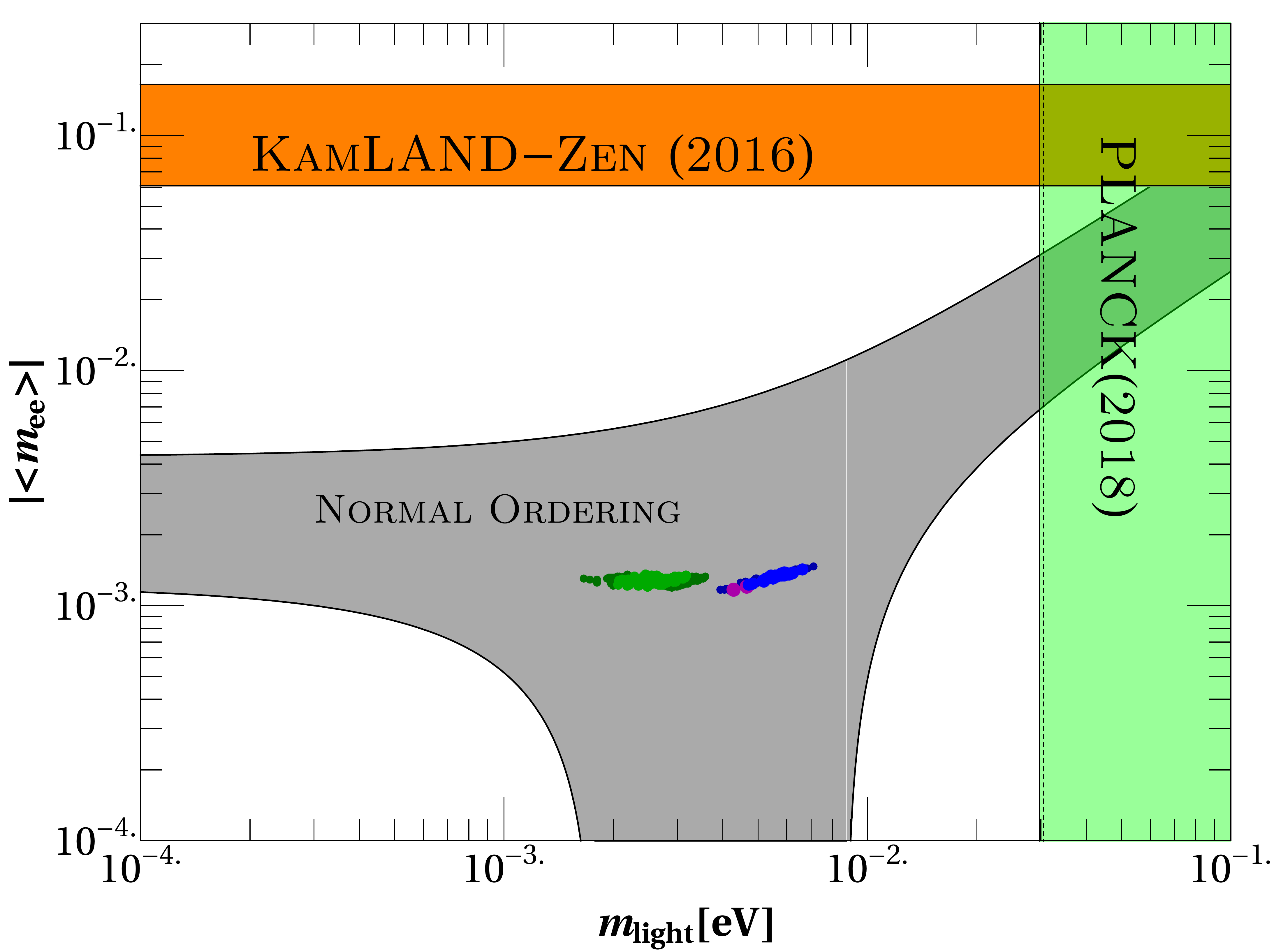}
\par\end{centering}
\caption{\footnotesize  The effective Majorana neutrino mass $|\langle m_{ee}\rangle |  $ vs the lightest neutrino mass $m_{light}$.  The prediction for the solutions B, C and D are shown by the color codes, which are same as the one used in Fig.~\ref{fig:Th23-CP}. Moreover, the latest upper bound on $|\langle m_{ee}\rangle |$ from the KamLAND-Zen collaboration are shown by the dark-yellow horizontal band. Also, the current results on the lightest neutrino mass is shown by the light green-vertical band from the \textit{Planck} Collaboration which gives $ \sum m_{\nu} < 0.12$ eV at the 95\% C.L. }
\label{fig:0nubb}
\end{figure}

Fig.~\ref{fig:0nubb} shows the prediction for $0\nu\beta\beta$ decay. For comparison, we first show the allowed $3\sigma$ parameter space in ($m_{light}-|\langle  m_{ee} \rangle|$)-plane using the latest global analysis of neutrinos oscillation data~\cite{deSalas:2017kay}, as shown by the  gray color. 
We proceed to compute the effective Majorana neutrino mass in Eq.~(\ref{0nubb}) for the allowed solutions,
namely for  B, C and D. The color code of the prediction  remains same
as the one used in Fig.~\ref{fig:Th23-CP}. The current upper bound on $|\langle  m_{ee} \rangle|$ comes from the KamLAND-Zen collaboration~\cite{KamLAND-Zen:2016pfg} which is read as $  |\langle  m_{ee} \rangle|  < (61 - 165)$ meV  at 90\% C.L. by taking into account the uncertainty in the estimation of the  nuclear matrix elements.
This is given by the dark-yellow horizontal band. The two black lines on this band corresponds to the uncertainty of the nuclear matrix element, $|M_{0\nu}|$ in Eq.~(\ref{eq:halflifedoublebeta}).  In addition, the light green-vertical band represents the bound on $m_{light}$ coming from the cosmological limit on the sum of neutrino masses provided by the Planck Collaboration, namely $ \sum m_{\nu} < 0.12$~eV at the 95\% C.L. ~\cite{Vagnozzi:2017ovm,Aghanim:2018eyx}.
Furthermore, as pointed out before, from the left-panel of Fig.~(\ref{fig:Th23-CP}) on can observe that the DUNE can rule out solution
D. This also has an impact for the prediction of $0\nu\beta\beta$. As a final remark, notice that the allowed solutions are compatible only with the normal neutrino mass ordering.
%
Finally, we proceed to discuss the possible lepton flavor violating (LFV) processes in the model. As we notice from the structure of the Yukawa couplings in Eq. (\ref{LY}), the charged leptons have non-diagonal couplings with the triplet scalar fields $\Delta_{L,R}$ and with the flavon fields $\eta$ and $\chi$. This may induce undesirably large tree level LFV processes such as $\mu \rightarrow e \gamma$ decay. Additionally, one-loop level diagrams with $W_{L,R}$ bosons also contribute to these processes. This processes may be used to delimit the parameter space of the LR gauge bosons and scalars. For example, the $\mu \rightarrow e \gamma$ branching ratio is given by~\cite{Cirigliano:2004mv} 
\begin{equation}
B\left(\mu\rightarrow e\gamma\right)=384\pi^{2}e^{2}\left(\left|A_{L}\right|^{2}+\left|A_{R}\right|^{2}\right) \;,
\end{equation}
where the functions $A_{L/R}$ are given in Appendix \ref{AppendixLFV} for completeness.
Using the Yukawa couplings in Eq.~(\ref{LY})
%
%
 we find $\textbf{BR}(\mu \rightarrow e \gamma) \leq 10^{-15} $, well below the experimental bound  $\textbf{BR}(\mu \rightarrow e \gamma)_{\text{exp}} \leq 4.2 \times 10^{-13}$ \cite{TheMEG:2016wtm} with the following LR parameters 
\begin{align}
 M_{W_R}&=6\text{ TeV}\;,     & M_{N_{1,2}}&= 5\text{ TeV}\;, & M_{\delta^{++}_{L/R}} &=4~ \text{TeV}\;, \nonumber \\ 
 M_{H^+} &=4~ \text{TeV}\; ,  & v_R &= 8 \text{ TeV}\;,       &  v_L &= 0.8 \text{ eV}\; , \nonumber \\
 v_\chi / \Lambda_F &=0.1\; , & v_\eta / v_\chi &=0.4\; . 
\end{align}
Notice that the benchmark values that have been considered here to calculate the branching ratio are in accordance with the neutrino oscillation results as discussed in Figs.~\ref{fig:Th23-CP},\ref{fig:0nubb}.
Moreover, one may also have contributions from the tree level flavon mediated processes. Given the hierarchy of scales of the model we expect them to be suppressed by the high scale of the flavon masses.
\section{Conclusions}
\label{sec:conclusions}

We have constructed a minimal left-right symmetric model with 
the addition of a flavor symmetry, the non-Abelian discrete
group $D_4$. We notice that besides the relations in the lepton 
Yukawas due to the left-right symmetry there are further correlations due to the additional family symmetry behind 
the theory. For this reason, there are a few free parameters left that can 
be written in terms of the 
leptonic observables, namely masses and their
mixing angles. This can be observed from the computation of the 
charged lepton mass matrix as well as the corresponding rotation matrix. 
The simplicity of the model leads to clear predictions for the neutrino sector.
Further, the model turns out to be compatible only with the normal
neutrino mass ordering and provides a correlation between the 
atmospheric angle $\theta_{23}$ and the leptonic CP-violating phase 
$\delta$. Given the possible solutions of the model there is one, namely the solution A is ruled
out by the current neutrino oscillation data. More importantly, due to the 
high potential of the DUNE experiment, which can improve the precision 
of $\theta_{23}$ and to probe $\delta$, it gives further restrictions
to the parameter space as shown in the right-panel in Fig.~(\ref{fig:Th23-CP}). Using this, the DUNE will be able to rule out two of the solutions, namely A and D.
We have also provided the prediction for neutrinoless double beta decay in terms
of the lightest neutrino mass for a mass range of $\sim (10^{-3}-10^{-2})$~eV, which we summarize in Fig.~(\ref{fig:0nubb}).
Finally, we have estimated the parameters of the LR sector where the LFV processes are under control. Branching ratio of $\mu \rightarrow e \gamma$ of the order of $\sim10^{-15} $ has been calculated, which is  well below the latest experimental bound.

\begin{acknowledgments}
The authors would like to thank Manuel E. Krauss, Toby Opferkuch and Werner Porod for  collaboration  in  the  early  stages  of  this  project.
This work is supported by the  German-Mexican  research  collaboration grant SP 778/4-1 (DFG) and 278017 (CONACYT),  CONACYT CB-2017-2018/A1-S-13051 (M\'exico) and DGAPA-PAPIIT IN107118 .  NN is supported by the postdoctoral fellowship program DGAPA-UNAM. 
The work of C.B. was also supported by the UCN grant ``Neutrino mass generation and BSM'' No. 20190803029 
C.B. would like to thank IFUNAM, Universidad de Colima-DCPIHP and Instituto de F\'isica Corpuscular (CSIC) for the hospitality 
while part of this work was carried out. RF and LMGDLV are supported by CONACYT.
\end{acknowledgments}

\begin{appendix}
\section{Basics of $D_4$ group}
\label{sec:d4multiplication}

The dihedral group $D_4$ is a non-Abelian group of order eight
and contains five irreducible representations (irreps), denoted as
$\bf{1},\,\bf{1^{'}},\,\bf{1^{''}},\,\bf{1^{'''}}$ and $\Rep{2}$, respectively. The 
two group generators $\rm A$ and $\rm B$ are
chosen as \cite{Adulpravitchai:2008yp}
\begin{equation}
\label{eq:generators}
\rm A =\left(\begin{array}{cc} 
                            i & 0 \\
                            0 & -i 
          \end{array}\right) \;\;\; \mbox{and} \;\;\; \rm 
		B=\left(\begin{array}{cc} 
                                       0 & 1 \\
                                       1 & 0 
                  \end{array}\right)  \;.
\end{equation}
For irreps $\Rep{2}$, they satisfy the following relations
\begin{equation}\label{eq:genrelations}
\mathrm{A}^{4} ={\mathbb I} \;, \;\;\; \rm B^2=\mathbb{I}, ~ \mbox{and}
\;\;\; \rm ABA=B \;.  
\end{equation}
where $\mathbb I$ is identity matrix.

The multiplication rules for the 1-dimensional irreps are the following
\begin{equation}\nonumber
\MoreRep{1}{a} \times \bf{1}= \MoreRep{1}{a} \; , \;\; 
\bf{1} \times \MoreRep{1}{a}= \MoreRep{1}{a} \;\; 
\; , \;\;
\bf{1^{'}} \times \bf{1^{''}}= \bf{1^{'''}} \; , \;\;
\bf{1^{'}} \times \bf{1^{'''}}= \bf{1^{''}} \;\; \mbox{and} \;\;
\bf{1^{''}} \times \bf{1^{'''}}= \bf{1^{'}} \; .
\end{equation}
For $(s_1,s_2,s_3,s_4) \sim ({\bf1^{}},{\bf1^{'}},{\bf1^{''}},{\bf1^{'''}})$ and $(x_1,x_2)^{T} \sim \Rep{2}$ we find
\begin{equation}\nonumber
\left( \begin{array}{c} s_1 x_1 \\ s_1 x_2
\end{array} \right) \sim \Rep{2} \;\; , \;\;\;
\left( \begin{array}{c} s_2 x_1 \\ -s_2 x_2
\end{array} \right) \sim \Rep{2} \;\; , \;\;\;
\left( \begin{array}{c} s_3 x_2 \\ s_3 x_1
\end{array} \right) \sim \Rep{2} \;\;\; \mbox{and} \;\;\;
\left( \begin{array}{c} s_4 x_2 \\ -s_4 x_1
\end{array} \right) \sim \Rep{2} \;\; .
\end{equation}
The product of a two-dimensional irreps $\Rep{2} \times \Rep{2}$ 
decomposes into the four singlets. Taking, for instance, $(x_1,x_2)^{T}\sim \Rep{2}$ 
and $(y_1, y_2)^{T}\sim \Rep{2}$ one finds
\begin{equation}
\nonumber
x_1 y_2 + x_2 y_1 \sim {\bf1} \;\; , \;\;\;
x_1 y_2 - x_2 y_1 \sim {\bf1^{'}} \;\; , \;\;\;
x_1 y_1 + x_2 y_2 \sim {\bf1^{''}} \;\;\; \mbox{and} \;\;\;
x_1 y_1 - x_2 y_2 \sim {\bf1^{'''}} \;\; .
\end{equation}

Now we show here the explicit form of Eq.~(\ref{LY}) using the $D_4$ product rules. We express the Lagrangian as
\begin{eqnarray}
\label{LYD4}
\mathcal{L}_Y
& \supset & \bar{\ell}_{L_{1}}\left(y_{1}\frac{\chi}{\Lambda_F}\Phi+\tilde{y}_{1}\frac{\chi}{\Lambda_F}\tilde{\Phi}\right)\ell_{R_{2}} + \bar{\ell}_{L_{2}}\left(y_{1}\frac{\chi}{\Lambda_F}\Phi+\tilde{y}_{1}\frac{\chi}{\Lambda_F}\tilde{\Phi}\right)\ell_{R_{1}} \nonumber \\
& + & \bar{\ell}_{L_{1}}\left(y_{2}\frac{\eta_2}{\Lambda_F}\Phi+\tilde{y}_{2}\frac{\eta_2}{\Lambda_F}\tilde{\Phi}\right)\ell_{R_{3}} + \bar{\ell}_{L_{2}}\left(y_{2}\frac{\eta_1}{\Lambda_F}\Phi+\tilde{y}_{2}\frac{\eta_1}{\Lambda_F}\tilde{\Phi}\right)\ell_{R_{3}} \notag\\
 & + & \bar{\ell}_{L_{3}}\left(y_{3}\frac{\eta_1}{\Lambda_F}\Phi+\tilde{y}_{3}\frac{\eta_1}{\Lambda_F}\tilde{\Phi}\right)\ell_{R_{2}} + \bar{\ell}_{L_{3}}\left(y_{3}\frac{\eta_2}{\Lambda_F}\Phi+\tilde{y}_{3}\frac{\eta_2}{\Lambda_F}\tilde{\Phi}\right)\ell_{R_{1}} \notag\\
 & + &\bar{\ell}_{L_{3}}\left(y_{4}\frac{\chi}{\Lambda_F}\Phi+\tilde{y}_{4}\frac{\chi}{\Lambda_F}\tilde{\Phi}\right)\ell_{R_{3}}\notag\\
 & + & \frac{Y_{L_{1}}}{2}\ell_{L_{1}}^{T}C\left(i\sigma_{2}\right)\Delta_{L}\ell_{L_{2}} + \frac{Y_{L_{1}}}{2}\ell_{L_{2}}^{T}C\left(i\sigma_{2}\right)\Delta_{L}\ell_{L_{1}} \notag\\
 & + &\frac{Y_{L_{2}}}{2}\ell_{L_{3}}^{T}C\left(i\sigma_{2}\right)\Delta_{L}\ell_{L_{3}} \notag\\
 & + & \frac{Y_{R_{1}}}{2}\ell_{R_1}^{T}C\left(i\sigma_{2}\right)\Delta_{R}\ell_{R_{2}} +  \frac{Y_{R_{1}}}{2}\ell_{R_2}^{T}C\left(i\sigma_{2}\right)\Delta_{R}\ell_{R_{1}} \notag\\
 & + & \frac{Y_{R_
 {2}}}{2}\ell_{R_{3}}^{T}C\left(i\sigma_{2}\right)\Delta_{R}\ell_{R_{3}}+\text{h.c.} 
\end{eqnarray}

The mass matrices for the charged-leptons (see Eq. (\ref{MMCL})) and neutrinos (see Eq. (\ref{MMnu})) are obtained from the Lagrangian \ref{LYD4}  after spontaneous symmetry breaking of the  gauge group.
%


We also show here that if one consider $\ell_{L_2}$, and $ \ell_{R_2}$ to be the $D_4$ singlets, 
the following Lagrangian can be obtained
\begin{eqnarray}
\label{LYD4-2}
\mathcal{L}_Y
& \supset & \bar{\ell}_{L_{1}}\left(y_{1}\frac{\chi}{\Lambda_F}\Phi+\tilde{y}_{1}\frac{\chi}{\Lambda_F}\tilde{\Phi}\right)\ell_{R_{3}} + \bar{\ell}_{L_{3}}\left(y_{1}\frac{\chi}{\Lambda_F}\Phi+\tilde{y}_{1}\frac{\chi}{\Lambda_F}\tilde{\Phi}\right)\ell_{R_{1}} \nonumber\\
& + & \bar{\ell}_{L_{1}}\left(y_{2}\frac{\eta_2}{\Lambda_F}\Phi+\tilde{y}_{2}\frac{\eta_2}{\Lambda_F}\tilde{\Phi}\right)\ell_{R_{2}} + \bar{\ell}_{L_{3}}\left(y_{2}\frac{\eta_1}{\Lambda_F}\Phi+\tilde{y}_{2}\frac{\eta_1}{\Lambda_F}\tilde{\Phi}\right)\ell_{R_{2}} \notag\\
 & + & \bar{\ell}_{L_{2}}\left(y_{3}\frac{\eta_1}{\Lambda_F}\Phi+\tilde{y}_{3}\frac{\eta_1}{\Lambda_F}\tilde{\Phi}\right)\ell_{R_{3}} + \bar{\ell}_{L_{2}}\left(y_{3}\frac{\eta_2}{\Lambda_F}\Phi+\tilde{y}_{3}\frac{\eta_2}{\Lambda_F}\tilde{\Phi}\right)\ell_{R_{1}} \notag\\
 & + &\bar{\ell}_{L_{2}}\left(y_{4}\frac{\chi}{\Lambda_F}\Phi+\tilde{y}_{4}\frac{\chi}{\Lambda_F}\tilde{\Phi}\right)\ell_{R_{2}}\notag\\
 & + & \frac{Y_{L_{1}}}{2}\ell_{L_{1}}^{T}C\left(i\sigma_{2}\right)\Delta_{L}\ell_{L_{3}} + \frac{Y_{L_{1}}}{2}\ell_{L_{3}}^{T}C\left(i\sigma_{2}\right)\Delta_{L}\ell_{L_{1}} \notag\\
 & + &\frac{Y_{L_{2}}}{2}\ell_{L_{2}}^{T}C\left(i\sigma_{2}\right)\Delta_{L}\ell_{L_{2}} \notag\\
 & + & \frac{Y_{R_{1}}}{2}\ell_{R_1}^{T}C\left(i\sigma_{2}\right)\Delta_{R}\ell_{R_{3}} +  \frac{Y_{R_{1}}}{2}\ell_{R_3}^{T}C\left(i\sigma_{2}\right)\Delta_{R}\ell_{R_{1}} \notag\\
 & + & \frac{Y_{R_{2}}}{2}\ell_{R_{2}}^{T}C\left(i\sigma_{2}\right)\Delta_{R}\ell_{R_{2}}+\text{h.c.}
\end{eqnarray} 
After the SSB of the gauge group, we obtain the charged-lepton mass matrix $M^{(2)}_{\ell}$ from Eq.~(\ref{LYD4-2}), whereas using the Dirac and Majorana neutrino mass matrices in Eq.~(\ref{MMnu}) one obtains the light neutrino mass matrix $M^{(2)}_{\nu}$. 
These can be read as
\begin{equation}
M^{(2)}_{\ell}   =\left(\begin{array}{ccc}
0 & 0 & a_{\ell}\\
0 & c_{\ell} & b_{\ell} \\
a^*_{\ell} & b^*_{\ell} & 0
\end{array}\right) \;, ~~~
M^{(2)}_{\nu} =\left(\begin{array}{ccc}
0 & 0 & a_{\nu}\\
0 & c_{\nu} & b_{\nu}\\
a_{\nu} & b_{\nu} & d_{\nu}
\end{array}\right)\;,
\end{equation}
where $a_{\nu}$, $b_{\nu}$, $c_{\nu}$, and $d_{\nu}$ are complex entries.

Similarly, if we choose  $\ell_{L_1}$,  and $\ell_{R_1}$ to be the $D_4$ singlets, then we find the following leptonic mass matrices
\begin{align}
\label{eq:ChargedLepton3}
M^{(3)}_{\ell} & =\left(\begin{array}{ccc}
c_{\ell} & b_{\ell} & 0\\
b^*_{\ell} & 0 & a_{\ell} \\
0 & a^*_{\ell} & 0
\end{array}\right) \;, ~~~
M^{(3)}_{\nu}  =\left(\begin{array}{ccc}
c_{\nu} & b_{\nu} & 0\\
b_{\nu} & d_{\nu} & a_{\nu}\\
0 & a_{\nu} & 0
\end{array}\right)\;.
\end{align}
One can also show that 
\begin{align}
\label{eq:ChargedLepton3}
M^{(2)}_{\ell} &  = R_{32} M_{\ell} R^T_{32},~~~  M^{(2)}_{\nu} = R_{32} M_{\nu} R^T_{32}  \;, \\
\label{eq:ChargedLepton4}
M^{(3)}_{\ell} &  = R_{31} M_{\ell} R^T_{31},~~~  M^{(3)}_{\nu} = R_{31} M_{\nu} R^T_{31} \;.
\end{align}
Here, $R_{31},$ and $ R_{32}$ are the permutation matrices  and are defined as 
\begin{equation}
R_{31} = \left(\begin{array}{ccc}
0 & 0 & 1\\
0 & 1 & 0\\
1 & 0 & 0
\end{array}\right), ~~~
R_{32} = \left(\begin{array}{ccc}
1 & 0 & 0\\
0 & 0 & 1\\
0 & 1 & 0
\end{array}\right) \;.
\end{equation}
It can be observed form Eq.~(\ref{eq:ChargedLepton3}) that the matrix $ R_{32}O_{\ell}$ diagonalizes  $M^{(2)}_{\ell}$ in analogous to Eqs.~(\ref{MCLorth}, \ref{eq:ChargedLepton1}), whereas $ R_{32}U_{\nu}$ helps to diagonalize  $M^{(2)}_{\nu}$ (see Eqs.~(\ref{eq:MnuTilde}, \ref{eq:leptonmixing})). Therefore, the leptonic mixing matrix  remains same as given by Eq.~(\ref{eq:pmns_sol}). Note that a similar conclusion also holds true for Eq.~(\ref{eq:ChargedLepton4}).

\section{$\mu \rightarrow e \gamma$ decay in left-right models }
\label{AppendixLFV}
In LR models the LFV processes are induced at tree level by the scalar triplets and at the loop level by the charged gauge bosons. Following the notation in \cite{Cirigliano:2004mv}, the branching ratio for $\mu \rightarrow e \gamma$ decay can be written as
\begin{equation}
B\left(\mu\rightarrow e\gamma\right)=384\pi^{2}e^{2}\left(\left|A_{L}\right|^{2}+\left|A_{R}\right|^{2}\right) \;,
\end{equation}
where 
\begin{equation}
A_{L}=\frac{1}{16\pi^{2}}\sum_{n=heavy}\left(K_{R}^{\dagger}\right)_{en}\left(K_{R}\right)_{n\mu}\left[\frac{M_{W_{L}}^{2}}{M_{W_{R}}^{2}}S_{3}\left(x_{n}\right)-\frac{x_{n}}{3}\frac{M_{W_{L}}^{2}}{M_{\delta_{R}^{++}}^{2}}\right] \;,
\end{equation}
\begin{equation}
A_{R}=\frac{1}{16\pi^{2}}\sum_{n=heavy}\left(K_{R}^{\dagger}\right)_{en}\left(K_{R}^{\dagger}\right)_{n\mu}x_{n}\left[-\frac{1}{3}\frac{M_{W_{L}}^{2}}{M_{\delta_{L}^{++}}^{2}}-\frac{1}{24}\frac{M_{W_{L}}^{2}}{M_{H_{1}^{+}}^{2}}\right] \;.
\end{equation}
Also,
\begin{equation}
x_{n}=\left(\frac{M_{n}}{M_{W_{R}}}\right)^{2},
\end{equation}
\begin{equation}
S_{3}\left(x\right)=-\frac{x\left(1+2x\right)}{8\left(1-x\right)^{2}}+\frac{3x^{2}}{4\left(1-x\right)^{2}}\left(S_{4}\left(x\right)+1\right),
\end{equation}
\begin{equation}
S_{4}\left(x\right)=\frac{x}{\left(1-x\right)^{2}}\left(1-x+\ln x\right),
\end{equation}
\begin{equation}
 K_{R}= (V_R^\nu)^\dagger V_R^l \;,
\end{equation}
where $V_R^{\nu / l }$ are the right handed neutrino/charged lepton mixing matrices respectively. We have neglected the small mixing of the charged gauge bosons, given the parameters of the LR gauge theory that were used in this prescription.

\end{appendix}
 
\bibliography{reference_new}

\end{document}